\documentclass[a4paper,12pt]{article}
\usepackage{amsmath, amsgen,amstext,amsbsy,amsopn,amsthm, amssymb, amscd}

\usepackage{graphicx, epstopdf}

\usepackage{tikz}

\usepackage{latexsym}

\newcommand{\Ibb}[1]{ {\rm I\ifmmode\mkern -3.6mu\else\kern -.2em\fi#1}}
\newcommand{\ibb}[1]{\leavevmode\hbox{\kern.3em\vrule
     height 1.2ex depth -.3ex width .2pt\kern-.3em\rm#1}}

\newtheorem{thm}{Theorem}

\theoremstyle{definition}

\newcommand{\be}{\begin{equation}}
\newcommand{\ee}{\end{equation}}
\newcommand{\beq}{\begin{equation}}
\newcommand{\eeq}{\end{equation}}
\newcommand{\bea}{\begin{eqnarray}}
\newcommand{\eea}{\end{eqnarray}}
\newcommand{\beqa}{\begin{eqnarray}}
\newcommand{\eeqa}{\end{eqnarray}}

\def\mfr#1/#2{\hbox{$\frac{{#1} }{ {#2}}$}}

\newcommand{\simt}{\mathrel{\rlap{\hbox{$\sim$}}\raise.9ex\hbox{{\tiny
$\,$T}}}}
\newcommand{\sima}{\mathrel{\rlap{\hbox{$\sim$}}\raise.95ex\hbox{{\tiny
$\,$A}}}}

\begin{document}
\title{The entropy concept for non-equilibrium states}
\author{Elliott H. Lieb${}^{1}$ and Jakob Yngvason${}^{2,3}$\\
\normalsize\it ${}^{1}$ Depts. of Mathematics and Physics, Princeton University, Princeton, NJ08542, USA\\
\normalsize\it ${}^{2}$ Faculty of Physics, University of Vienna, Austria.\\
\normalsize\it ${}^{3}$ Erwin Schr{\"o}dinger Institute for Mathematical Physics, Vienna, Austria.}
\date{}
\maketitle 
\begin{abstract}
In earlier work we presented a foundation for the Second Law of Classical Thermodynamics in terms of the Entropy Principle. More precisely, we provided an empirically accessible axiomatic derivation of an entropy function defined on all equilibrium states of all systems that has the appropriate additivity and scaling properties and whose increase is a necessary and sufficient condition for an adiabatic process between two states to be possible. Here, after a brief review of this approach, we address the question of defining entropy for non-equilibrium states. Our conclusion is that it is generally not possible to find a unique entropy that has all relevant physical properties. We do show, however, that one can define two entropy functions, called $S_-$ and $S_+$, which, taken together, delimit the range of adiabatic processes that can occur between non-equilibrium states. The concept of {\it comparability} of states with respect to adiabatic changes 
plays an important role in our reasoning.
\end{abstract}

\section{Introduction}

It is commonly held that entropy increases with time. While entropy is fairly unambiguously well defined for equilibrium states, a good part of the matter in the 
universe, if not most of it, is not in an equilibrium state. It does not have a well defined entropy
as one measures it using equilibrium concepts such as Carnot cycles, for example, but if one does 
not know precisely what entropy is for non-equilibrium systems, the notion of increase 
cannot be properly quantified.  

Several definitions of entropy for non-equilibrium states have been proposed in the literature. (See, e.g.,  \cite{LJC} for a review of these matters and \cite{ST} for a discussion of steady-state thermodynamics.)
These definitions do not necessarily fulfill the main requirement of entropy, however, which,
according to our view, is that entropy is a state function that allows us 
to determine {\it precisely} which changes are possible and which are not under well 
defined conditions.  Given the magnitude of this challenge, we do not
mean to criticize the heroic efforts of many scientists to define non-equilibrium entropy and utilize it
for practical calculations, but we would like  to point out here
some of the problems connected with defining 
entropy in non-equilibrium situations.

Our starting point is the basic empirical fact that under ``adiabatic conditions'' certain changes of the 
{\it equilibrium} states of thermodynamical systems are possible and some are not.
The {\it Second Law of Thermodynamics} (at least for us) is the assertion that
the possible state changes are characterized by the increase 
(non-decrease) of an (essentially) unique state function,  called 
{\it Entropy}, that is extensive, and additive on subsystems.


The Second Law is one of the few really
fundamental physical laws. It is independent of models and its consequences are far reaching.  
Hence it deserves a simple and solid logical foundation!
An approach to the foundational issues was developed by us in several papers in 1998-2003 \cite{LY1}-\cite{LY4}. 
We emphasize that, contrary to possible first impression, our approach is not abstract but  is
based, in principle, on experimentally determined facts.  We also emphasize that 
our approach is independent of concepts from statistical mechanics and model making. 
This point of view has recently been taken up and even applied to engineering thermodynamics  in the textbook
by  A.\ Thess \cite{T}.

We can summarize the contents of the present paper as follows. We begin, in the next section,  with a very brief review of our approach to the meaning and existence of 
entropy for equilibrium systems.  An important concept 
 is the {\it adiabatic comparability} ({\it comparability} for short) of states with respect to the basic relation of {\it adiabatic accessibility} (to be explained in the next section). This property, that is usually taken for granted in traditional approaches, often without saying so, means that for any two states $X$ and $Y$ of the same chemical composition there exists an adiabatic process that either takes $X$ to $Y$ or the other way around. If one assumes this {\it a priori}, then the existence and uniqueness of entropy follows in our approach quickly from some very simple and physically plausible axioms. However, it is argued in \cite{LY1}-\cite{LY4} that this comparability is, in fact, a highly nontrivial property that needs justification. The mathematically most sophisticated part of this earlier work, and its analytical backbone, is the establishment of comparability starting from some simpler physical assumptions that include convex combinations of states, continuity property of generalized pressure, and assumptions about thermal contact.
 
In the following section we discuss the possibilities for extending the definition of entropy to non-equilibrium states. The concept of comparability will again play an important role. In fact, we shall argue that it may not be possible in general to define one unique entropy for non-equilibrium states that fulfills all the roles of entropy for equilibrium states. 
Instead one has to expect a whole range of entropies lying between two extremes, which we call $S_-$ and $S_+$. Only when comparability holds do these two state functions coincide and we have a unique entropy. Comparability for non-equilibrium states, however,  is an even less trivial property than for equilibrium states and can certainly not be expected in general.

Another point that comes into play and is far from trivial is {\it reproducibility} of states. In fact, it is hard to talk about the properties of states that occur only once in the span of the universe, but that is often the case for non-equilibrium states. For this reason one must be circumspect about definitions that may look good on paper but can not be implemented in fact.


\section{The entropy of classical equilibrium thermodynamics}  

This section gives a summary of the main findings of \cite{LY1}-\cite{LY4}. We consider
thermodynamical systems, that can be simple or compound, and have equilibrium states denoted by $X,X'\dots$. These states are collected in state spaces $\Gamma,\Gamma', \dots$.  The {\it composition} (also called `product')
$(X,X')$ of a state $X\in\Gamma$ and $X'\in\Gamma'$, which means simply considering the two states jointly but without a physical interaction between them,  is just a point in the cartesian product $\Gamma\times\Gamma'$. There is also the concept of a {\it scaled copy} $\lambda X\in\lambda\Gamma$ of a state $X\in \Gamma$ with a real number $\lambda>0$. This means that extensive properties like energy, volume etc.\ are scaled by $\lambda$ while intensive properties like pressure, temperature, etc.\ are not changed. Composition and scaling are supposed to satisfy some obvious algebraic rules.

To begin with, state spaces are just sets, and no more structure is needed for the `elementary' part of our approach. However, for the further development and in particular the derivation of adiabatic comparability, we assume that the state spaces are open convex subsets of $\mathbb R^N$ for some $N\geq 2$ (depending on the state space). {\it Simple systems}, that are the building blocks for composite systems, have a distinguished coordinate, the energy $U$, and $N-1$ work coordinates, denoted collectively by $V$. Often, $V$ is just the volume.

A central concept in our approach (as in \cite{C}, \cite{L},  \cite{B}, \cite{FJ}, \cite{G}) is the relation of {\it adiabatic accessibility}. Its 
operational definition (inspired by Planck's formulation of the Second 
Law, see  \cite{P}. p. 89) is as follows:

{\it A state $Y$ is  adiabatically accessible from a state $X$, in
symbols $X\prec Y$ (read: ``$X$ precedes $Y$"),
if it is possible to change  the state from $X$ to
$Y$ in such a way that the final effect on the surroundings is that a 
weight may have risen or fallen.}\footnote{Such processes are called {\it work processes} in \cite{GB}.}

It is important to note that the process taking $X$ to $Y$ need not be ``quasi-static", in fact, it can be arbitrarily violent.
   
 The following definitions and notations will be applied:  
    If $X\prec Y$ {\it or} $Y\prec X$ we say that  $X$ and $Y$ are  {\it adiabatically
comparable} (or {\it comparable} for short). If $ X\prec Y$ {\it but}
 $Y\not\prec X$ we write $
X\prec \prec Y
$ (read: ``$X$ strictly precedes $Y$"), and if both
$X\prec Y$ {\it and} $Y\prec X$ hold we write 
$X\sima Y$ and say that $X$ and $Y$ are {\it adiabatically equivalent}.
\subsection{The entropy principle}

We can now state the {\bf Second Law (Entropy Principle)}:
\medskip
   
    {\it There is a function called {\bf Entropy}, 
defined on all states and denoted by $S$, such that the following holds:\\ \\
\noindent a) {\bf Characterization of adiabatic accessibility:} For two states $X$ and $Y$ with the same `matter content' \footnote{In \cite{LY1}-\cite{LY4} the `matter content'  is measured by the scaling parameters of some basic simple systems. Physically, one can think of  the amounts of the various chemical ingredients of the system.}
\beq X \prec Y\  \mbox{\rm if and only if} \  S(X) \leq S(Y).\eeq\\
\noindent b) {{\bf Additivity and extensivity:}} For compositions and scaled copies of states we have }
\beq S(X, X') = S(X) + S(X')\quad{\rm and}\quad S(\lambda X)=\lambda S(X).\eeq

      \noindent {\bf Remarks}\\ 
      
\noindent      1. The scaling relation in (2) says that the entropy doubles when the
size of the system is doubled, but this linearity is not a triviality.
It need not hold for nonequilibrium entropy, where non-linear effects
might come into play.

The additivity in (2) is one of the remarkable facts about entropy
(and one of the most difficult to try to prove if there ever is a
mathematical proof of the second law from assumptions about dynamics).
The states $X$ and $X'$ can be states of two different systems, yet
(2) says that the amount by which one system can reduce its entropy in
an adiabatic interaction of the two systems is precisely offset by the
minimum amount by which the other system is forced to raise its entropy.\\

       \noindent 2.  It is noteworthy that the {\it mere existence} of entropy satisfying the fundamental relation
     \beq\label{basicrel} dS=\frac 1T dU+\frac PT dV-\sum_i\, \frac{\mu_i}T dn_i\eeq
where $T=(\partial S/\partial U)^{-1}$ is the absolute temperature, $P=T(\partial S/\partial V)$ the (generalized) pressure and the $\mu_i=T(\partial S/\partial n_i)$ the chemical potentials of the constituents with mole numbers $n_i$ in a mixture, leads to surprising connections between quantities that at first sight look unrelated, for instance\footnote{The first equation is the relation between the velocity of sound and heat capacities in a dilute gas, and the second is the Clausius-Clapeyron equation with $P_0(T)$ the pressure at the coexistence curve between two phases, $\Delta h$ the specific latent heat and $\Delta v$ the specific volume change at the phase transition. The last equation is the van 't Hoff relation between the equilibrium constant $K(T)$ of a chemical reaction and the heat of reaction, $\Delta H$.}
\beq \frac{ m\,v_{\rm sound}^2}{RT}=\frac{c_P}{c_V}, \quad \frac{dP_0}{dT}=\frac{\Delta h}{T\Delta v} ,\quad \frac d{dT}\ln K(T)=\frac {(\Delta H)^2}{RT^2}.\eeq
 
 \noindent       3. Another consequence of the existence of entropy is a formula, due to Max Planck (\cite{P}, pp.\ 134-135), that relates 
  an arbitrary {empirical temperature} scale $\Theta$ to the absolute temperature scale $T$:
  
     \begin{equation}\label{5}
   T(\Theta)=T_0 \exp
   \left(\,\,
   \int_{\Theta_0}^\Theta{\frac{\left(\frac{\partial P}{\partial \Theta'}\right)_{\!V}}{P+\left(\frac{\partial U}{\partial V}\right)_{\!\Theta'}}\;d\Theta'}
  \right)
  \end{equation}
  It is remarkable that the integral on the right-hand side depends only on the temperature although the terms in the integrand depend in general also on the volume, but this follows from the fact that \eqref{basicrel} is a total differential.\\

\noindent    4. The entropy also determines the {\it maximum work} that can be obtained from a system in an environment with temperature $T_0$:
    \beq\Phi_{X_0}(X)=(U-U_0)-T_0(S-S_0)\eeq
    where $X$ is the initial state with energy $U$ and entropy $S$, and $X_0$ is the final state with energy $U_0$ and entropy $S_0$.
    (This quantity is also called {\it availability} or {\it exergy}.)\\
    
   The {\bf main questions} that were addressed in \cite{LY1}-\cite{LY4} are:
   
    \begin{itemize}
  \item[Q1] Which properties of the relation $\prec$ ensure existence and (essential)
uniqueness of entropy?
  \item[Q2] Can these properties be derived from simple physical premises?
\item[Q3] Which further properties of entropy follow from the premises?
\end{itemize}      
To answer Q1 the following conditions on $\prec$  were identified in \cite{LY1}-\cite{LY4}:

    \begin{itemize} 
 \item[A1]  {\it Reflexivity\/}:  $X \sima X$.

 \item[A2]  {\it Transitivity:\/} If $X \prec Y$ and $Y \prec Z$, then $X 
\prec Z$.

 \item[A3] {\it Consistency\/}: If $X \prec X^\prime$ and $Y 
\prec Y^\prime$, then $(X,Y) \prec
(X^\prime, Y^\prime)$.

 \item[A4] {\it Scaling Invariance\/}: If $\lambda > 0$ and
$X,Y \in \Gamma$ with $X \prec Y$,  then
$\lambda X \prec \lambda Y$.

 \item[A5]  {\it Splitting and Recombination\/}: 
$X \sima ((1-\lambda) X, \lambda X).$

 \item[A6]  {\it Stability\/}: If
$(X, {\varepsilon} Z_0) \prec (Y, {\varepsilon} Z_1)$
for some $Z_0$, $Z_1$ and a sequence of $\varepsilon$'s tending to zero, then
$X \prec Y.$
    \end{itemize}
    
These six conditions are all 
highly plausible if $\prec$ is interpreted as the relation of adiabatic 
accessibility in the sense of the operational definition.
They are, however, not sufficient to ensure the existence of 
an entropy that characterizes the relation on compound systems made of 
scaled copies of $\Gamma$. A further property is needed:

\begin{itemize}\item[CP]{ \it Comparison Property for scaled products of a state space $\Gamma$}: 
Any two states in $(1-\lambda)\Gamma\times \lambda \Gamma$ are adiabatically
comparable, 
for all $0\leq \lambda\leq 1$.
\end{itemize}
 {}   
\subsection{Uniqueness and the basic construction of entropy for equilibrium states}
If one assumes  (CP) together with (A1)-(A6), it is a simple matter to prove that  the entropy on a state space $\Gamma$ is {\it uniquely determined}, up to an affine change of scale, provided an entropy function exists. The proof goes as follows.

We first pick two reference points $X_0\prec\prec X_1$ in $\Gamma$. ({Recall that $X_0\prec\prec X_1$ means that $X_0\prec X_1$, but $X_1 \prec X_0$ does {\it not} hold.} If there are no such points, then all points in $\Gamma$ are adiabatically equivalent and  the entropy must be a constant.)  Suppose  $X$ is an arbitrary state with $X_0 \prec X\prec  X_1$. (If $X\prec X_0$, or $X_1\prec X$, we interchange the roles of $X$ and $X_0$, or $X_1$ and $X$, respectively.)  For any entropy function $S$ we have $S(X_0)<S(X_1)$ and $S(X_0)\leq S(X)\leq S(X_1)$ so there is a unique number $\lambda$ between 0 and 1 such that
\beq S(X)=(1-\lambda)S(X_{0})+\lambda S(X_{1}).\eeq
By the assumed properties of entropy this is  {equivalent} to
\beq\label{8} X\sima ((1-\lambda) X_{0},\lambda 
X_{1}).\eeq
Any other entropy function $S'$  also leads to \eqref{8} with $\lambda$ replaced by some  $\lambda'$. From the assumptions A1-A6 and $X_0\prec\prec X_1$ it is easy to prove
 (see \cite{LY2}, Lemma 2.2) 
 that 
\eqref{8} can hold for {\it at most} one $\lambda$, i.e., $\lambda=\lambda'$.
Hence the entropy is uniquely determined, up to the choice of the entropy of $X_0$ and $X_1$, i.e., up to an affine change of scale.


We now come to the {\it existence of entropy}.
From assumptions A1-A6 {and CP} one shows (see \cite{LY4}, Eqs. (8.13)-(8.20)) that 
\beq\sup\{\lambda\,:\, ((1-\lambda)X_0,\lambda X_1)\prec X\}=\inf\ \{\lambda\,:\, X\prec ((1-\lambda)X_0,\lambda X_1)\},
\eeq
and, denoting this number by $\lambda^*$,  \beq\label{11} X\sima ((1-\lambda^*) X_{0},\lambda^*
X_{1}).\eeq
With the choice 
\beq S(X_{0})=0\quad {\rm and}\quad S(X_{1})=1\eeq for some reference 
points $X_{0}\prec\prec X_{1}$, we now have an  {\bf explicit formula for the entropy}: 
\beq\label{entropydef}\boxed{S(X)=\sup\{\lambda\,:\, ((1-\lambda)X_0,\lambda X_1)\prec X\}}\eeq
or, equivalently, 
\beq\label{entropydef2}\boxed{S(X)=\inf\ \{\lambda\,:\, X\prec ((1-\lambda)X_0,\lambda X_1)\}.}\eeq
These formulas use  only the relation $\prec$ and make neither appeal to Carnot cycles nor to statistical mechanics.\footnote{If $X_1\prec\prec  X$, then $((1-\lambda)X_0,\lambda X_1)\prec X$ has the meaning $\lambda X_1\prec((\lambda-1)X_0, X)$ and the entropy is $>1$. Likewise, the entropy is $<0$ if $X\prec\prec  X_0$. See \cite{LY2}, Remark 2, pp.\ 27--28. }

       {}

A change of reference points is clearly equivalent to an affine change of scale for $S$.
Thus the main conclusion so far is:

\begin{thm} [{\bf Existence and uniqueness of entropy, given CP}]
 The existence and  uniqueness (up to a choice of scale) of entropy on $\Gamma$ is {\bf equivalent} to  
assumptions A1-A6  together with the comparison property, CP.\end{thm}

\noindent {\bf Remarks}

\noindent 1. The uniqueness is very important. It means that any other definition leading to an entropy function satisfying the requirements of the second law as stated above, is {\it identical} (up to a scale transformation) to the entropy defined by Eq. \eqref{entropydef}. In order to {\it measure} $S$ it is not necessary to resort to Eq. \eqref{entropydef} or \eqref{entropydef2}, although it is in principle possible to do so. (Note that the use of $\sup$ and $\inf$ is not a mathematical abstraction but merely reflects the fact that in reality measurements are never perfect). Instead, one can  use any method, like the standard practice when preparing steam tables, for instance,  namely measuring heat capacities, compressibilities etc., using Eq.\ \eqref{5} to convert empirical temperatures to absolute temperatures, and then integrating Eq.\ \eqref{basicrel} along an arbitrary path from a reference state to the state whose entropy is to be determined.\\

\noindent 2. The {\it  comparison property} (CP)  plays a central role in our reasoning and it is appropriate to make some comments on it. First, we emphasize that,
in order to derive  \eqref{11}, comparability of all states in
$(1-\lambda)\Gamma\times \lambda \Gamma$,  and not only of those in $\Gamma$, is essential. Previous authors have also noted the importance of comparability. In the seminal work of R. Giles \cite{G} it appears as the requirement that if $X$, $Y$ and $Z$ are any states (possibly of composite systems) such that $X\prec Z$ and $Y\prec Z$, then $X$ and $Y$ are comparable. The same conclusion is assumed if $Z\prec X$ and $Z\prec Y$. Similar requirements were made earlier by Landsberg \cite{L},  Buchdahl \cite{B}, and Falk and Jung \cite{FJ}. These assumptions imply that states fall into equivalence classes such that comparability holds within each class. That comparability is nontrivial, even for equilibrium states, can be seen from the example of systems that have only energy as a coordinate (`thermometers')  and where only `rubbing' and thermal equilibration are allowed as adiabatic operations. For the composite of two such systems CP is violated and the entropy is not unique. See \cite{LY2}, Fig. 7, p. 65, and also Sect.\ 3.4  and Fig. 3 below.

While the references above are only concerned with equilibrium states, the authors of \cite{GB} require comparability as part of their second law, even for non-equilibrium states. We shall comment further on this issue in the next section.\\

We do not want to adopt CP as an axiom because we do not find it physically compelling. Our preference is to derive it from some more immediate assumptions. Consequently, an essential part of the analysis in \cite{LY1}-\cite{LY4}, and, in fact,  mathematically the most complex one, is a {\it derivation} of CP from additional assumptions about {\it simple systems}
which are the basic building blocks of thermodynamics. At the same time one makes contact with the traditional concepts of thermodynamics such as pressure and temperature.

As already mentioned, the states of simple  systems are described by one 
{energy coordinate} $U$ (the { First Law} enters here) and one or more  work coordinates, like the 
volume $V$.  The key assumptions we make are:
\begin{itemize} \item[{$\bullet$}] The possibility of 
forming, by means of an adiabatic process,  {\it convex combinations}  of states of simple systems with 
respect to the energy $U$ and the work coordinate(s)  $V$.\vskip-.5cm
 \item[{$\bullet$}] The existence of 
{ at least one irreversible adiabatic state change}, starting from 
{ any} given state. 
 \end{itemize}
{} 
       {}
      \begin{itemize}
      \item Unique supporting planes for the convex sets $\mathcal A_X=\{Y:\; X\prec Y\}$ (`forward sectors') and a regularity property for their slope (the generalized pressure).
                  \end{itemize}
      From these assumptions one derives (\cite{LY2}, Theorems 3.6 and 3.7)\\
        
\begin{thm}[{\bf Comparability of states for simple systems}]
{\it Any two  states $X$, $Y$, of a simple system are comparable, i.e., either $X\prec Y$ or $Y\prec X$.
 Moreover, $X\sima Y$ if and only if  $Y$ lies on the boundary of $\mathcal A_X$, or equivalently,   $X$ lies on the boundary of $\mathcal A_Y$.}
 \end{thm}
 
 This theorem, however, is not enough because to define $S$ by means of the formulas 
 \eqref{entropydef}-\eqref{entropydef2} we need more, namely {\it comparability for all states in
 $(1-\lambda)\Gamma\times \lambda \Gamma$, not only in $\Gamma$!}\\
 
 The additional concept needed is
 \begin{itemize}
\item {Thermal equilibrium} between simple systems, in particular the {Zeroth Law}.\end{itemize}
In essence, this allows us to {\it make one simple system out of the compound system  $(1-\lambda)\Gamma\times \lambda \Gamma$} so that the  previous analysis can be applied to it, eventually leading to comparability for all states in the compound system.  See \cite{LY2} Section 4.\\

The final outcome of the analysis is (cf.\ \cite{LY2}, Theorems 4.8 and 2.9):\\

\begin{thm}[{\bf Entropy for equilibrium states}]
The comparison property is valid for arbitrary scaled products of 
simple systems. Hence the relation among states in such state spaces 
is characterized by an {addi\-tive and extensive entropy}, $S$. \\
The entropy is {unique} up to an overall multiplicative constant and one 
additive constant for each `basic' simple system.\\
Moreover, the entropy is a {concave} function of the energy and work 
coordinates, and it is {nowhere locally constant}.\end{thm}

To include  {mixing processes} and {chemical reactions} as well, the entropy constants for different mixtures of given ingredients, and also of  compounds of the chemical elements, have to be chosen in a {consistent} way. In our approach it can be proved, {without invoking idealized `semipermeable membranes'},  that the entropy scales of the various substances can be shifted in such a way
that $X\prec Y$ always implies $S(X)\leq S(Y)$. The converse, i.e., that
 $S(X)\leq S(Y)$ implies $X\prec Y$ provided $X$ and $Y$
have the same chemical composition, cannot be guaranteed without further assumptions, however. These matters are discussed in \cite{LY2}, Sect.\ 6.

 \section{Non-equilibrium states} 
 
There exist many variants of non-equilibrium thermodynamics. A concise overview is given in \cite{LJC} where the following approaches are discussed, among others: Classical Irreversible Thermodynamics, Extended Irreversible Thermodynamics, Finite Time Thermodynamics,  Theories with Internal Variables, Rational Thermodynamics, Mesoscopic Thermodynamic Descriptions.  Most of these formalisms focus on states close to equilibrium. Aspects of steady state thermodynamics are thoroughly discussed in \cite{ST}.

A further point to note is that the role of entropy in non-equilibrium thermodynamics is considerably less prominent than in equilibrium situations. Equilibrium entropy is a thermodynamic potential when given as a function of its natural variables $U$ and $V$, i.e., it encodes {\it all} equilibrium thermodynamic properties of the system. For a description of non-equilibrium phenomena, on the other hand,  more input than the entropy alone is needed. 

It is a meaningful question, nevertheless,  to ask to what extent an entropy can be defined for non-equilibrium states, preserving as much as possible of the useful properties of equilibrium entropy. To formulate and discuss this question precisely, we consider a system with a space $\Gamma$ of equilibrium states that is embedded as a subset in some larger space 
 $\widehat\Gamma$ of  {non-equilibrium} states. We emphasize that $\widehat \Gamma$ need not contain 
 {\it all} non-equilibrium states that the system might in principle possess, but only a part that is relevant for the problems under consideration.  A natural requirement is that states in $\widehat \Gamma$ are {\it reproducible}. It is not clear to us that the entropy of an exploding bomb, for instance, is a meaningful concept (although the energy might be).
 
 Another point to keep in mind is that a non-equilibrium state is, typically, either time dependent, 
 or it is not isolated from its environment, as in the case of a non-equilibrium steady state that has to
be connected to  reservoirs that cause fluxes of heat or electric current to flow through  it.  
 
 \subsection{Entropies for non-equilibrium states}
 
We assume that a relation $\prec$ is defined on $\widehat\Gamma$ such that its restriction to the equilibrium state space $\Gamma$ is characterized by an entropy function $S$, as discussed in Section 2. The physical meaning of $\prec$ on $\hat \Gamma$ is supposed to be the same as before, i.e., $X\prec Y$ means that $Y$ can be reached from $X$ by a process that in the end leaves no traces in the surroundings except that a weight may have been raised or lowered. As discussed in the previous section the assumption that the restriction $\prec$  to $\Gamma$ is characterized by an entropy function is equivalent to assuming that this restriction satisfies conditions A1-A6 together with CP. For the non-equilibrium states in $\widehat\Gamma$ it is {\it not} natural to require A4 (scaling) and A5 (splitting), but we shall assume the following:
 \begin{itemize}
\item[N1] The relation $\prec$  on $\widehat \Gamma$  satisfies our assumptions A1 (Reflexivity), A2 (Transitivity), A3 (Consistency)\footnote{Compound states have the same meaning as in the equilibrium situation, i.e., we consider two copies of the system and one state of each copy side by side.}
and A6 (Stability). 

\item[N2] For every $X\in \widehat \Gamma$ there are $X', X''\in \Gamma$ such that $X'\prec X\prec X''$. \end{itemize}

The meaning of the second requirement is that every non-equilibrium state in $\widehat\Gamma$ can be generated from an equilibrium state in $\Gamma$ by an adiabatic process, and that every non-equilibrium state can be brought to equilibrium by such a process (e.g., by letting the non-equilibrium state relax spontaneously to equilibrium). We consider this to 
be very natural, physically.\\

The {\bf basic question} we now ask is: \textit{What can be said about possible extensions of $S$ to functions $\widehat S$ on $\widehat\Gamma$  that are monotone with respect to $\prec$, i.e., that satisfy $\widehat S(X)\leq \widehat S(Y)$ if $X\prec Y$, and, if $X\in \Gamma$, then $\widehat S(X)=S(X)$ as well?}\\
 
Our answer involves the following {\it two} functions:\\

For $X\in\widehat\Gamma$ define

\begin{equation}\label{14}\boxed{\phantom{\int}S_-(X):=\sup\{S(X')\,:\, X'\in \Gamma, X'\prec X\}\,\, }\end{equation}

\begin{equation}\label{15}\boxed{\phantom{\int}S_+(X):=\inf\ \{S(X'')\,:\, X''\in \Gamma, X\prec X''\}\,\, }
\end{equation}

Thus  $S_-$ measures how large the entropy can be of an equilibrium state out of which $X$ is created by an adiabatic process, and $S_+$ measures
how small the entropy  of an equilibrium state can be into which $X$ equilibrates by an adiabatic process.

The essential properties of these functions are collected in the following Proposition.  In words, it says the following: Both $S_-$ and  $S_+$ take only finite values and increase or remain unchanged under adiabatic state changes.  A sufficient condition for $Y$ to be adiabatically accessible from $X$ is that $S_+(X)\leq S_-(Y)$.  While neither of the functions are necessarily additive, $S_-$ is at least superadditive and $S_+$ subadditive  (see Eq. (18) below).  The entropy is unique if and only if $S_-=S_+$, because any function, that is monotonously increasing or unchanged with respect to the relation of adiabatic accessibility and coincides with $S$ on $\Gamma$, lies between $S_-$ and $S_+$. The unique entropy is then additive, by (18).\\

\noindent {\bf Proposition 1} ({\bf  Properties of ${S_\pm}$}).   
\begin{itemize}
\item[(a)] $-\infty< S_\pm(X)<+\infty$ for all $X\in\widehat\Gamma$.

\item[(b)] $S_\pm(X)=S(X)$ for $X\in\Gamma$, and $S_-(X)\leq S_+(X)$, 
 for all $X\in\widehat \Gamma$.
\item[(c)] The $\sup$ and $\inf$ in the definition of $ S_\pm$ are attained for some $X', X''\in \Gamma$ with $X'\prec X\prec X''$.

\item[(d)] \beq\label{17}\boxed{\hbox{$X\prec Y$ implies $S_-(X)\leq S_-(Y)$ and  $S_+(X)\leq S_+(Y)$.}}\eeq
(See Figure 1.)  

\item[(e)] \beq\label{18}\boxed{\hbox{If $S_+(X)\leq S_-(Y)$, then $X\prec Y$.}}\eeq
(See Figure 2.)


\item[(f)] Under composition, $S_-$ is superadditive and $S_+$ subadditive, i.e., 
\begin{equation}\boxed{S_-(X_1)+S_-(X_2)\leq S_-(X_1,X_2)\leq S_+(X_1,X_2)\leq S_+(X_1)+S_+(X_2)}\end{equation}
\item[(g)] If $\widehat 
S$ is any function on $\widehat \Gamma$ that coincides with $S$ on $\Gamma$  and is such that  $X\prec Y$ implies $\widehat S(X)\leq \widehat S(Y)$, then
\beq S_-(X)\leq \widehat S(X) \leq S_+(X).\eeq
\end{itemize}

\begin{proof} (a) and (b) follow immediately from the assumptions on $\prec$ and the properties of $S$, namely $X'\prec X\prec X''$ implies $S(X')\leq S(X'')$.

\smallskip
(c) Since the entropy is concave on $\Gamma$ and in particularly continuous, it takes all values between $S(X')$ and $S(X'')$ for any two states $X'$ and $X''$  in $\Gamma$ with $X'\prec X''$. Hence, by N2 and the definition of $S_-$, there is an $X_0'\in\Gamma$ with $S(X'_0)=S_-(X)$, and we claim that $X_0'\prec X$. Indeed, by the definition of $S_-$ there is, for every $\varepsilon>0$,
an $X_\varepsilon'\in\Gamma$ with $X_\varepsilon'\prec X$ and
$0\leq S(X'_0)-S(X_\varepsilon')\leq \varepsilon$. We can pick $Z_0\prec\prec Z_1\in \Gamma$ and $0\leq \delta(\epsilon)$ such that $\delta(\varepsilon)\to 0$ for $\varepsilon\to 0$ and $S(X'_0)+\delta S(Z_0)=S(X_\varepsilon')+\delta S(Z_1)$. Then
$$(X'_0,\delta Z_0)\sim (X_\varepsilon',\delta Z_1)\prec (X,\delta Z_1).$$
Hence $X'_0\prec X$ by the stability assumption A6.  In the same way one shows that the infimum defining $S_+(X)$ is attained.

\smallskip

(d) If $X\prec Y$ and $X'\prec X$, then $X'\prec Y$, so $S_-(X)\leq S_-(Y)$. Likewise, $X\prec Y$ and $Y\prec Y''$ implies $X\prec Y''$, so also $S_+(X)\leq S_+(Y)$.

\smallskip

(e) If $S_+(X)\leq S_-(Y)$ then there exists $X''$ and $Y'$ with $X\prec X''$, $Y'\prec Y$ and $X''\prec Y'$. But then $X\prec Y$ by transitivity.

\smallskip

(f) By (c) there exist $X_i',X_i''$, $i=1,2\in \Gamma$  such that $S_-(X_i)=S(X_i')$, $S_+(X_i)=S(X_i'')$ and
$X_i'\prec X_i\prec X_i''$. From the additivity of the equilibrium entropy $S$ and
$$(X_1',X_2')\prec(X_1,X_2)\prec (X_1'',X_2'')$$
we obtain
\begin{multline}S_-(X_1)+S_-(X_2)=S(X_1')+S(X_2')=S(X_1',X_2')\\ \leq S_-(X_1,X_2) \leq S_+(X_1,X_2) \leq S(X_1'',X_2'')\\=
S(X_1'')+S(X_2'')=S_+(X_1)+S_+(X_2).\end{multline}

\smallskip

 {} 

(g) Let $X'\prec X\prec X''$ as in (c). Then
\begin{equation}S_-(X)=S(X')=\widehat S(X') \leq \widehat S(X)\\ \leq \widehat S(X'')=S(X'')=S_+(X).\end{equation}\end{proof}

The following Theorem clarifies the connection between adiabatic comparability  and  uniqueness of an extension of the equilibrium entropy to the non-equilibrium states.
(Recall that  two states, $X$ and $Y$ are called {comparable} w.r.t.\ the relation $\prec$ if either $X\prec Y$, or $Y\prec X$ holds.) Particularly noteworthy is the equivalence of (i), (iii) and (vi) below, that may be summarized as follows:  {\it A non-equilibrium entropy, characterizing the relation $\prec$, exists if and only if every 
non-equilibrium state is adiabatically equivalent to some equilibrium state.}


   {} 

\begin{thm}[{\bf Comparability and uniqueness of entropy}]
The following are equivalent:
\begin{itemize}
\item[(i)] There exists a {\it unique} $\widehat S$ extending $S$ such that $X\prec Y$ implies $\widehat S(X)\leq \widehat S(Y)$.
 
\item[(ii)] $S_-(X)=S_+(X)$ for all $X\in\widehat\Gamma$.

\item[(iiii)] There exists a (necessarily unique!) $\widehat S$ extending $S$ such that $\widehat S(X)\leq \widehat S(Y)$ implies  $X\prec Y$.

\item[(iv)] Every $X\in\widehat\Gamma$ is comparable with every $Y\in\widehat\Gamma$, i.e., the Comparison Property is valid  on $\widehat\Gamma$.

\item[(v)] Every $X\in\widehat\Gamma$ is comparable with every $Z\in\Gamma$.

\item[(vi)] Every $X\in\widehat\Gamma$ is adiabatically equivalent to some $Z\in\Gamma$.
\end{itemize}
\end{thm}
   {} 
\begin{proof} That (i) is equivalent to (ii) follows from (d) und (g) in Proposition 1. 
Moreover, (ii) implies (iii) by (e).  The implications (iii) $\rightarrow$ (iv) $\rightarrow$ (v) are obvious. 

(v) $\rightarrow$ (ii): If $X'$ and $X''$ are as in Prop.1 (c), and 
$S(X')=S_-(X)<S_+(X)=S(X'')$  there exists a $Z\in\Gamma$ with $S(X')<S(Z)<S(X'')$. If (v) holds, then either $Z\prec X$ or $X\prec Z$. The first possibility contradicts the definition  of $S_-$ and the latter definition  of $S_+$. Hence $S_-=S_+$, so (v) implies (ii).  

It is clear that (vi) $\rightarrow$ (v) because CP holds on $\Gamma$.

 Finally, (ii) $\rightarrow$ (vi): Since $X', X''\in\Gamma$, and $S(X')=S(X'')$ (by (ii)), we know that $X'\sima X''$, because the entropy $S$ characterizes the relation $\prec$ on $\Gamma$ by assumption. Now  $X'\prec X\prec X''$, so $X\sima X'\sima X''$.
  \end{proof}

   {} 
  
\subsection{Maximum work}

Assume now that  a `thermal reservoir' with temperature $T_0$ is given. Such a reservoir can be regarded an idealization of a simple system without work coordinates that is so large that an energy change has no appreciable effect on its temperature (defined, as usual, to be the inverse of the derivative of the entropy with respect to the energy).
An energy change $\Delta U_{\rm res}$ and a corresponding entropy change $\Delta S_{\rm res}$ of the reservoir are thus connected by
\beq\label{21}\Delta U_{\rm res}=T_0 \Delta S_{\rm res}.\eeq
We denote by $\Phi_{X_0}(X)$ the maximum work that can be extracted from a state $X\in\widehat \Gamma$ with the aid of the reservoir if the system ends up in a state $X_0\in\Gamma$ with internal energy $U_0$ and entropy $S_0$. If $X\in\Gamma$, i.e., $X$ is an equilibrium state, then 
\beq \Phi_{X_0}(X)=(U-U_0)-T_0(S-S_0)\eeq
where $U$ is the internal energy of $X$ and $S$ its entropy. This follows as usual by considering the total entropy change of system plus reservoir, i.e., $S_0-S+\Delta S_{\rm res}$, which has to be $\geq 0$ by the second law for equilibrium states. 
The work extracted is $W=-(\Delta U_{\rm res}+U_0-U)$,
and using  \eqref{21} we obtain $W\leq (U-U_0)-T_0(S-S_0)$. Equality is reached if the process is reversible.
\\

For non-equilibrium states $X\in\widehat \Gamma$ the $\pm$ entropies defined in \eqref{14} and \eqref{15} give at least upper and lower bounds  for the maximum work:
\begin{equation}\label{24}
\boxed{ \phantom{\int}(U-U_0)-T_0(S_+-S_0)\leq \Phi_{X_0}(X)\leq (U-U_0)-T_0(S_--S_0). \,\, }\end{equation}
where $S_\pm$ denote the $\pm$ entropies of $X$. This can be seen as follows:\\

Consider a special process $X\rightarrow X''\rightarrow X_0$ where the first step is an adiabatic process and where
$X''$ and $X_0$ are equilibrium states.
Since, by definition,  there is no change to the rest of the universe in the first process other than the motion of a weight, 
conservation of energy tells us that 
the work  obtained in the step $X\rightarrow X''$ is $U-U''$. In the step $X''\rightarrow X_0$ the maximum work (by the standard reasonings for  equilibrium states in $\Gamma$, see above) is $(U''-U_0)-T_0(S(X'')-S_0)$. Altogether
\begin{multline*}\Phi_{X_0}(X)\geq U-U''+(U''-U_0)-T_0(S_+-S_0)=(U-U_0)-T_0(S_+-S_0)\end{multline*} where we have used that $S(X'')=S_+$ for $X''$ as in Prop.\ 1(c).
An analogous reasoning applied to $X'\rightarrow X\rightarrow X_0$ (with $X'$ as in Prop.\ 1(c)) gives
$$\Phi_{X_0}(X')=U'-U_0-T_0(S_--S_0)\geq U'-U+\Phi_{X_0}(X)$$ and hence $\Phi_{X_0}(X)\leq (U-U_0)-T_0(S_--S_0)$.\\

   {} 

\subsubsection{Definition of entropy through maximum work}

In their influential textbook on engineering thermodynamics \cite{GB}, E. Gyftopolous and G.\ P. Beretta (see also \cite{BZ})
take the concept of maximum work as a basis for their definition of entropy. Paraphrasing their definition in our notation, 
they assume the maximum work $\Phi_{X_0}(X)$ to be a measurable quantity for {\it arbitrary} states $X$ (equilibrium as well as non-equilibrium) and define an entropy $S_{\rm GB}(X)$  through the formula
{\beq\label{25}\Phi_{X_0}(X) =(U-U_0)-T_0(S_{\rm GB}(X)-S_0).\eeq}
From Eq.\ \eqref{24} it is clear that
\begin{equation}S_-(X)\leq S_{\rm GB}(X)\leq S_+(X).\end{equation}
This follows also from Prop.\ 1(g) because
\begin{equation} X\prec Y \text{ implies } S_{\rm GB}(X)\leq S_{\rm GB}(Y)\end{equation}
that can be seen by considering  the process $X\rightarrow Y\rightarrow X_0$, obtaining
\begin{multline*}U-U_0-T_0(S_{\rm GB}(X)-S_0)=\Phi_{X_0}(X)\geq U-U(Y)+\Phi_{X_0}(Y)\\ =
U-U(Y)+U(Y)-U_0-T_0(S_{\rm GB}(Y)-S_0)\\=U-U_0-T_0(S_{\rm GB}(Y)-S_0).
\end{multline*}

   {}

The Gyftopolous-Beretta entropy is therefore {\it one} possible choice of a function that is monotone w.r.t.\ $\prec$.  According to our analysis it characterizes the relation $\prec$ if and only if the Comparison Property is valid on the whole state space $\hat \Gamma$ (as GB {\it assume} as part of their Second Law; see also Assumption 2 in \cite{BZ}), in which case {\it all} entropies on $\hat \Gamma$ extending $S$ coincide. In particular, the GB approach via maximum work leads to the {\it same} equilibrium entropy as the approach of Section 2.

As we have already stated, however,  and shall discuss further below, we consider it implausible to assume adiabatic comparability for {\it general} non-equilibrium states. If CP does not hold on $\hat\Gamma$, the entropy $S_{\rm GB}$ may depend in a nontrivial way on the choice of the thermal reservoir and the final state $X_0$\footnote{The proof of independence in \cite{GB} 
and \cite{BZ} uses the assumption that any state can be transformed into an equilibrium state by a reversible work process, which amounts to assuming property (vi) in Theorem 4 above.}.  This means that the availability for a different final state $\tilde X_0$ and different reservoir with temperature $\tilde T_0$ is {\it not} necessarily given by the formula $\Phi_{\tilde X_0}(X) =(U-\tilde U_0)-\tilde T_0(S_{\rm GB}(X)-\tilde S_0)$ if $S_{\rm GB}(X)$ is defined by means of \eqref{25}.

On the other hand, each of the entropies $S_\pm$, which are also monotonic w.r.t.\ $\prec$ by Proposition 1 (b), is unique up to a scale transformation, since these entropies are defined in terms of the equilibrium entropy on $\Gamma$ which has this uniqueness property. The inequalities \eqref{24} hold for all $T_0$ and $X_0$, irrespective of comparability.

\subsection{Why adiabatic comparability is implausible in general}
  
 According to Theorem 4 the Comparison Property on $\hat\Gamma$, and hence the (essential) uniqueness of entropy,  is {\it equivalent} to the statement that {\it every non-equilibrium state $X\in\hat\Gamma$ is adiabatically equivalent to some equilibrium state $Z\in\Gamma$}. While there are idealized situations when such comparability can be conceived, it seems to be a highly implausible property in general. The problem can already be expected to arise close to equilibrium as we now discuss. 
  
  Consider first the `benign' case where ``Classical Irreversible Thermodynamics" (CIT) (see \cite{LJC}, Ch.\ 2) can be considered an adequate approximation. The states in $\hat \Gamma$ are here described by local values of equilibrium parameters like temperature, pressure and matter density. In particular, one can define a local entropy density by using the equilibrium equation of state, and by subsequently integrating  this entropy density over the volume of the system one obtains a global entropy. An equilibrium state in $\Gamma$ of the same system with the same entropy is, to a good approximation,  adiabatically equivalent to the non-equilibrium state. This can be seen by dividing the system into cells such that each cell is approximately in equilibrium and regarding the collection of cells as a composite equilibrium system for which the comparison property holds by the analysis described in Section 2.
  
The situation changes, however, when CIT is not adequate and the {\it fluxes} have to be considered as independent variables, as in ``Extended irreversible Thermodynamics"  (EIT)
(cf.\ \cite{LJC}, Ch. 7). In this situation, also the local temperature has to be replaced by a different variable (cf. \cite{LJC}, Section 7.1.2).\footnote{On the microscopic level, a Maxwell-Boltzmann distribution for the velocities of the molecules in a small volume element gets shifted to a different distribution.} A phenomenological ``extended entropy", depending explicitly on the heat flux can be considered and even computed in some simple cases. It has the property of increasing monotonously in time when heat conduction is described by Cattaneo's model \cite{Cat} with a hyperbolic heat transport equation rather than the parabolic Fourier's law.\footnote{Fourier's Law is $\mathbf q=-\lambda \nabla T$ where $\mathbf q$ is the heat flux, $\lambda$ the heat conductivity and $T$ the temperature. Cattaneo's Law is $\tau\partial \mathbf q/\partial t=-(\mathbf q+\lambda \nabla T)$ where $\tau$ is the time constant of heat flux relaxation.} The classical entropy, in contrast, may oscillate (see Fig. 7.2 in \cite{LJC}), and does therefore  not comply with the second law. Also, the argument above for establishing adiabatic equivalence with equilibrium states no longer applies. Although we do not have a rigorous proof, we consider it highly implausible that a state that is significantly influenced by the flux can be adiabatically equivalent to an equilibrium state, where no flux is present, for this would mean that turning the flux on or off could be done reversibly. Unless this can be done, however, CP does {\it not} hold on $\widehat\Gamma$ and there is no unique entropy.

If one moves further away from equilibrium, not even EIT may apply and CP becomes even less plausible. In extreme cases like an exploding bomb one may even question whether it is meaningful to talk about entropy as a state function at all, because the highly complex situation just after the explosion can not be described by reproducible macroscopic variables. 

For systems with reproducible states, the entropies $S_\pm$ are at least well defined and in principle measurable, although it may not be easy to do so in practice. They provide bounds on the possible adiabatic state changes in the system and the maximum work that can be extracted from the system in a given state and a given environment. The difference 
\beq \Delta S(X):=S_+(X)-S_-(X),\eeq
which is unique up to a universal multiplicative factor, can also be considered as a measure of the deviation of $X$ from equilibrium.

 \subsection{A Toy Example}

To elucidate the concepts and issues discussed above we may consider a simple toy example. The system consists of two  identical pieces of copper that are glued together by a thin layer of finite heat conductivity. We regard the state of the system as uniquely specified by the energies, or equivalently, the temperatures $T_1$ and $T_2$  of the two copper pieces that are assumed to have constant heat capacity. The layer between them is considered to be so thin that its energy can be ignored. Mathematically,  the state space $\hat \Gamma$ of this system is thus $\mathbb R_+^2$ with coordinates $(T_1,T_2)$ and the equilibrium state space $\Gamma$ is the diagonal, $T_1=T_2$.  

We assume to begin with that the relation
$\prec$ is defined by the following `restricted' adiabatic operations: 
\begin{itemize}
\item Increasing the energy of each of the copper pieces by rubbing.
\item Heat conduction between the pieces through the connecting layer obeying Fourier's law.
\end{itemize}
The forward sector $\mathcal A_X=\{Y:\, X\prec Y\}$ of $X=(T_1,T_2)$ then consists of all points that can be obtained by rubbing, starting from any point on the line segment between $(T_1,T_2)$ and the equilibrium point $(\hbox{$\frac12$}(T_1+T_2), \hbox{$\frac12$}(T_1+T_2))$ (See Figure 3). 

As equilibrium entropy  we take $S(T,T)=\log T$. The points $X'$ and $X''$ of Prop.\ 1(c) are
$$X'=(\min\{T_1,T_2\},\min\{T_1,T_2\})\qquad X''=(\hbox{$\frac12$}(T_1+T_2), \hbox{$\frac12$}(T_1+T_2))$$(see Figure 3)
and accordingly
$$S_-(T_1,T_2)=\min \{\log T_1,\log T_2\} \qquad S_+(T_1,T_2)=\log(\hbox{$\frac12$}(T_1+T_2)).$$

If we {\it extend} the relation $\prec$ defined above by allowing the copper pieces to be temporarily taken apart and using them as thermal reservoirs between which a Carnot machine can run  to equilibrate the temperatures reversibly, then the previous forward sector will be extended and is now characterized by the unique extension of $S$ to
$$\hat S(T_1,T_2)=\hbox{$\frac12$}(\log T_1+\log T_2).$$
This is precisely the 'benign' situation referred to at the beginning of Section 3.3 where CIT applies.

  If the parts are unbreakably linked together, however, the situation is different. An irreversible heat flux between the two parts during the adiabatic state change is then unavoidable. If the heat conduction is governed by Cattaneo's rather than Fourier's law it is necessary to introduce the heat flux as a new independent variable and apply EIT as discussed in the last section.  The general objections against the CP and hence the existence of a unique entropy mentioned then apply. But even if we stay with Fourier's law and the two dimensional state space of the toy model, it is clear that  the extended forward sector, obtained by applying Carnot machines in addition to rubbing and equilibration, will depend on the relation between the heat conductivity of the separating layer between the parts and the heat conductivity
between the Carnot machine and the copper pieces. If the latter is finite, a gap between $S_-$ and $S_+$ will remain,  because equilibration of the temperatures by means of the Carnot machine requires a minimal nonzero time span, during which heat leaks irreversibly through the layer connecting the two pieces.

\section{Summary and Conclusions}
  
\quad  \ \  1. Under the stated general assumptions A1-A6 for equilibrium states, and  N1-N2 for non-equilibrium states, the possibility of defining a single, {\it unique} entropy, monotone 
  with respect to  the relation of adiabatic accessibility, 
is  {\it equivalent to the adiabatic comparability of states} (CP). \medskip

2. Comparability is a {highly nontrivial} property. Even in the equilibrium situation it requires additional axioms beyond A1-A6. \medskip

   {}
3. It is implausible that comparability holds for arbitrary non-equilibrium states. It might, however,  be established for 
\textit{restricted classes} of non-equilibrium states. In any case, a prerequisite for a useful definition of entropy is  that the states can be {\it uniquely identified} and that they are {\it reproducible}.  \medskip

4. Further insight into the question of comparability might be obtained from concrete models in which the relation
$\prec$ is defined by some dynamical laws.

\bigskip

\noindent{\small\textbf{Acknowledgments.}\\
Work partially supported by  U.S. National Science Foundation (grant
PHY 0965859; E.H.L.), the Simons Foundation (grant \# 230207; E.H.L), and the Austrian Science Fund
FWF (P-22929-N16; JY). We thank the Erwin Schr\"odinger Institute of the University of Vienna
for its hospitality and support.}

\newpage

\begin{center}
\begin{tikzpicture}[>=latex]
\draw[very thick, ->>](-6,0) --(-0.8,0);
\draw[very thick](-6,0) --(6,0);
\draw[very thick, ->>](-0.8,0) --(2.5,0);
\draw[very thick](2.5,0) --(6,0);
\draw[thick] (-4,0) -- (-3,2.5) -- (-2.5,0);

\draw[red, very thick, ->>] (-4,0) -- (-3.5,1.25);
\draw[red, very thick] (-3.5,1.25) -- (-3,2.5);

\draw[blue, very thick, ->>] (-3,2.5) -- (-2.75,1.25);
\draw[blue, very thick] (-2.75,1.25) -- (-2.5,0);

\draw[thick] (-3.5,0)
(1.5,5.5) -- (4,0);t
\draw[red, very thick, ->>] (-3.5,0) -- (-1,2.75);
\draw[red, very thick] (-1,2.75) --(1.5,5.5);
\draw[blue, very thick, ->>] (1.5,5.5) -- (2.75,2.75);
\draw[blue, very thick] (2.75,2.75) -- (4,0);
\draw[very thick,->>] (-3,2.5) -- (-0.75,4);
\draw[very thick] (-0.75,4) -- (1.5,5.5);
\node at (-3,2.5) {$\bullet$};
\node at (-4,0) {$\bullet$};
\node at (-2.5,0) {$\bullet$};
\node at (-3.5,0) {$\bullet$};
\node at (1.5,5.5) {$\bullet$};
\node at (1,0) {$\bullet$};
\node at (4,0) {$\bullet$};
\node at (-3,3) {$X$};
\node at (1.5,6) {$Y$};
\node at (6.5,0) {\large\large$\Gamma$};
\node at (4.5,4) {\large\large${\hat\Gamma}$};
\node at (-4,-.5) {$X'$};
\node at (-3.5,-.5) {$Y'$};
\node at (-2.5,-.5) {$X''$};
\node at (4,-.5) {$Y''$};
\node at (1,-.5) {$Z$};

\end{tikzpicture}.
\end{center}
\medskip
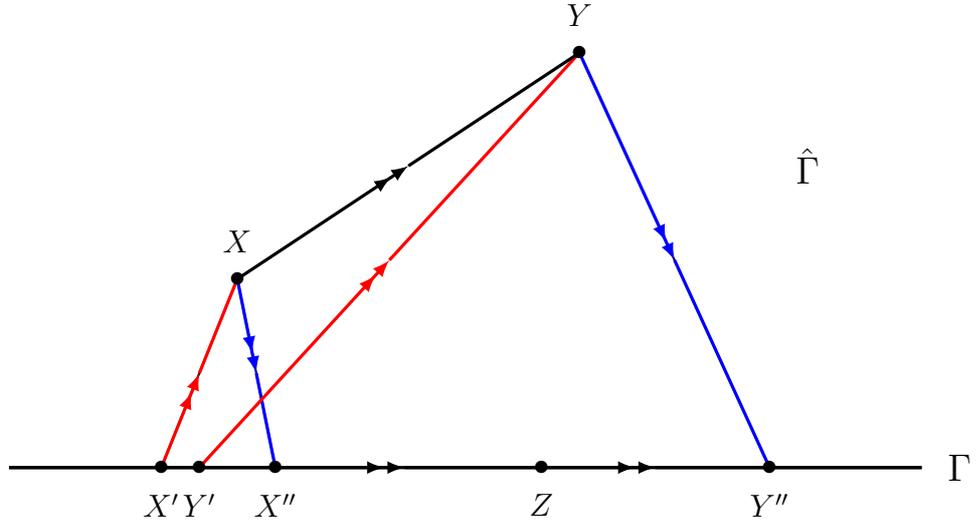
{\small Fig.\ 1.  The figure illustrates Eq.\ (\ref{17}) with $S_-(X)=S(X')$, $S_+(X)=S(X'')$,\\ $S_-(Y)=S(Y')$, $S_+(Y)=S(Y'')$. The arrows indicate adiabatic state changes. The state  $Z\in\Gamma$ with $Y'\prec \prec Z\prec\prec Y''$ is not adiabatically comparable with $Y$ (but it is adiabatically comparable with $X$ because $X\prec X''\prec Z$).}

\begin{center}
\begin{tikzpicture}[>=latex]
\draw[very thick](-6,0) --(-2.5,0);
\draw[very thick, ->>](-2.5,0) --(-1.25,0);
\draw[very thick](-1.25,0) --(4,0);
\draw[very thick](-2.5,0) --(4,0);



\draw[blue, very thick, ->>] (-3,2.5) -- (-2.75,1.25);
\draw[blue, very thick] (-2.75,1.25) -- (-2.5,0);

\draw[red, very thick, ->>] (0,0) -- (0.75,2.75);
\draw[red, very thick] (0.75,2.75) --(1.5,5.5);
\draw[very thick, dashed,->>] (-3,2.5) -- (-0.75,4);
\draw[very thick, dashed] (-0.75,4) -- (1.5,5.5);
\node at (-3,2.5) {$\bullet$};
\node at (-2.5,0) {$\bullet$};
\node at (0,0) {$\bullet$};
\node at (1.5,5.5) {$\bullet$};
\node at (-3,3) {$X$};
\node at (1.5,6) {$Y$};
\node at (4.5,0) {\large\large$\Gamma$};
\node at (4.5,4) {\large\large${\hat\Gamma}$};
\node at (0,-.5) {$Y'$};
\node at (-2.5,-.5) {$X''$};
\end{tikzpicture}.
\end{center}
\medskip
{\small Fig.\ 2.  Illustration of Eq.\ (\ref{18}) with $S_+(X)=S(X'')$, $S_-(Y)=S(Y')$.}
\newpage

\begin{center}
\begin{tikzpicture}[>=latex]
\draw[thick, ->](0,0) --(0,5);
\draw[thick, ->](0,0) --(5,0);

\fill[red!30!white] (5,2)--(2,2)--(1,3)--(1,5)--(5,5)--(5,2);
\draw[very thick](0,0) --(5,5);
\node at (1,3) {$\bullet$};
\draw[thick](1,3) --(1,5);
\draw[thick](1,3) --(2,2);
\draw[thick](2,2) --(5,2);
\draw[dashed](1,3) --(1,1);

\node at (3,4) {$\mathcal A_X$};

\node at (.7,3) {$X$};
\node at (1.2,.8) {$X'$};
\node at (2.2,1.7) {$X''$};
\node at (5.4,0) {$T_1$};
\node at (0,5.4) {$T_2$};
\node at (5.2,5.2) {$\Gamma$};

\end{tikzpicture}
\end{center}

\medskip
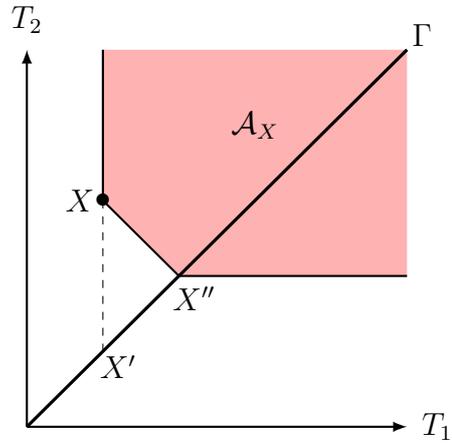
{\small Fig.\ 3.  Illustration of $\mathcal A_X$ and the points $X'$ and $X''$ in the
toy example in Sect. 3.4.}


\newpage

\begin{thebibliography}{99}
\bibitem{LJC} Lebon,  G., Jou, D. \& Casas-V\'azquez, J. 2008 {\it Understanding Non-equilibrium Thermodynamics: Foundations, Applications, Frontiers}.  Berlin, Heidelberg: Springer-Verlag.

\bibitem{ST} Sasa, Sh.-I. \& Tasaki, H. 2006 Steady state thermodynamics. {\it J.\ Stat.\ Phys.} 125, 125--224.


 \bibitem{LY1} Lieb, E. H. \& Yngvason, J. 1998 A guide to entropy and the second law of thermodynamics. {\it Notices of the Am. Math. Soc.} 
45, 571-581.

\bibitem{LY2} Lieb, E. H. \& Yngvason, J. 1999  The physics and mathematics of the second law of thermodynamics. {\it Phys. Rep.} 310, 1--96; Erratum 314, 669.

\bibitem{LY3} Lieb, E. H. \& Yngvason, J. 2000  A fresh look at entropy and the second law of thermodynamics, {\it 
Phys.\ Today}  53, Nr. 4, 32--37; Entropy Revisited, Gorilla and All, {\it 
Phys.\ Today} 53, Nr. 10, 12--13.

\bibitem{LY4} Lieb, E.H. \& Yngvason, J.  2003 The Entropy of Classical Thermodynamics, in {\it Entropy} (ed. A. Greven, G. Keller \& G. Warnecke), pp. 147--193. Princeton: Princeton University Press. 

\bibitem{T}  Thess, A. 2011 {\it  The Entropy Principle:  Thermodynamics for the Unsatisfied}. Berlin, Heidelberg: Springer-Verlag.

\bibitem{C} Carath\'eodory, C. 1909 Untersuchung \"uber die Grundlagen der Thermodynamik, {\it Math. Annalen}  67, 355--386

\bibitem{G} Giles, R. 1964 {\it Mathematical Foundations of Thermodynamics}. Oxford: Pergamon Press. 

\bibitem{L} Landsberg, P.T. 1956  Foundations of thermodynamics. {\it Rev. Mod. Phys.} 28, 363-392

\bibitem{B} Buchdahl, H. A. 1958  A formal treatment of the consequences of the second law of thermodynamics in Carath\'eodory's
formulation. {\it Zeits.f. Phys.} 152, 425-439.

 \bibitem{FJ} Falk, G.; Jung, H. 1959 Axiomatik der Thermodynamik, in {\it Handbuch der Physik} (ed. S. Fl\"ugge), vol III/2, pp. 199-175. Springer-Verlag.
       
\bibitem{P} Planck, M. 1945 {\it Treatise on Thermodynamics}. New York: Dover Publications. 


\bibitem{GB} Gyftopoulos, E.P. \& Beretta, G.P. 1991, 2nd ed. 2005 {\it Thermodynamics: foundations and applications}. New York: Dover Publications. 

\bibitem{BZ} Beretta, G.P. \& Zanchini, E. 2011 Rigorous and General Definition of Thermodynamic Entropy, in {\it Thermodynamics} (ed.  M. Tadashi), ISBN:978-953-307-544-0. InTech, 2011, Available from {\tt http://www.intechopen.com/books/thermodynamics/} pp. 24--50. 

\bibitem{Cat} Cattaneo, C. 1945 Sulla conduzione del calore. {\it Atti Seminario Mat. Fis. Univ. Modena}  3, 83-101.
 
\end{thebibliography}
\end{document}